\documentclass[runningheads]{llncs}
\usepackage[T1]{fontenc}
\usepackage{graphicx}
\usepackage{amsmath}
\usepackage{amssymb}
\usepackage{relsize}
\usepackage{multirow}
\usepackage{booktabs}  % For tables 
\usepackage{comment}  % To comment out a large blocks of text

\usepackage{algorithm}
\usepackage{algorithmic}

\usepackage{enumitem}

\newcommand{\be}{\begin{equation}}
\newcommand{\ee}{\end{equation}}

\DeclareMathOperator*{\argmin}{arg\,min}

\UseRawInputEncoding
 
\begin{document}
\title{Structural breaks detection and variable selection in dynamic linear regression via the Iterative Fused LASSO in high dimension}

\titlerunning{Structural break and variable selection}
% If the paper title is too long for the running head, you can set
% an abbreviated paper title here
%
\author{Angelo Milfont\inst{1}\orcidID{0009-0005-9925-5082} \and
Alvaro Veiga\inst{1}\orcidID{0000-0003-0200-6724} }
\authorrunning{A. Milfont, A. Veiga}
% First names are abbreviated in the running head.
% If there are more than two authors, 'et al.' is used.
%
\institute{Pontifical Catholic University of Rio de Janeiro, Rio de Janeiro, Brasil}
%\email{lncs@springer.com}\\
%\url{http://www.springer.com/gp/computer-science/lncs} \and
%ABC Institute, Rupert-Karls-University Heidelberg, Heidelberg, Germany\\
%\email{\{abc,lncs\}@uni-heidelberg.de}}
%
\maketitle              % typeset the header of the contribution

\begin{abstract}
We aim to develop a time series modeling methodology tailored to high-dimensional environments, addressing two critical challenges: variable selection from a large pool of candidates, and the detection of structural break points, where the model's parameters shift. This effort centers on formulating a least squares estimation problem with regularization constraints, drawing on techniques such as fused LASSO and adaptive LASSO, which are well-established in machine learning. Our primary achievement is the creation of an efficient algorithm capable of handling high-dimensional cases within practical time limits. By addressing these pivotal challenges, our methodology holds the potential for widespread adoption. To validate its effectiveness, we detail the iterative algorithm and benchmark its performance against the widely recognized Path Algorithm for Generalized LASSO. Comprehensive simulations and performance analyses highlight the algorithm's strengths. Additionally, we demonstrate the methodology's applicability and robustness through simulated case studies and a real-world example involving a fund under active management, where the number of shares of each asset at each time instant can change. These examples underscore the methodology's practical utility and potential impact across diverse high-dimensional settings.
\keywords{Dynamic Regression \and Lasso  \and Adaptive Lasso \and Fused Lasso \and Generalized Lasso \and Variable Selection \and High dimension}
\end{abstract}

\section{Introduction}
Time series models with exogenous variables are extensively applied across diverse fields such as economics, business, finance, environmental studies, and health. These models fulfill two primary objectives: analyzing the dynamics of a variable to identify influential factors, and forecasting the variable's future evolution. Despite their widespread use, the practical application of these models is hampered by major challenges: selecting relevant variables, and identifying structural breaks. 

The challenge of variable selection, often referred to as determining the "model structure," has traditionally been addressed through the sequential application of statistical tests \cite{Breaux1967}. While effective for small-scale problems (fewer than 100 variables), these methods become impractical in high-dimensional contexts where the number of candidate variables exceeds the available sample size \cite{Epprecht2021}. In such scenarios, alternative approaches from machine learning, particularly regularization techniques, have proven invaluable. These methods estimate coefficients by minimizing the squared error while incorporating penalties on the magnitude of the coefficients. Ridge Regression \cite{Hoerl1970} was among the first of these techniques, though it does not perform variable selection, as all coefficients remain non-zero. The introduction of LASSO \cite{tibshirani1996lasso} marked a significant breakthrough, enabling efficient handling of a large number of variables. This was later refined by methods such as the Adaptive LASSO \cite{Zou2006}, which improved performance in variable selection and model fitting.

Another critical issue in time series modeling is the identification of structural breaks, moments when the coefficients, or even the set of relevant variables, undergo significant changes, which  
%In finance, dynamic regression models used to replicate active investment fund strategies must account for corresponding changes in regression coefficients over time.
%The identification of structural breaks and the detection of change points in high dimension settings 
often arises in the financial domain of active fund management. In this context, accurate detection is crucial for effective performance evaluation, risk management, and investment oversight. Failure to identify such shifts can result in the continued use of outdated models, misestimation of risk, and poor capital allocation decisions. Conversely, timely detection allows investors to adjust strategies, monitor potential style drifts, and reassess fund manager effectiveness, thereby enhancing portfolio resilience and aligning investment decisions with current market conditions.

Extensive econometric research addresses structural breaks, from seminal works such as \cite{Chow1960} to more recent studies like \cite{Perron1989}. However, as with variable selection, these methods are largely based on repeated statistical tests, limiting their applicability to low-dimensional settings. In high-dimensional contexts, the literature remains underdeveloped, particularly in addressing the simultaneous challenges of variable selection and structural breaks. The Fused LASSO \cite{tibshirani2005fused}, which implements a "trend filter," offers a partial solution by modeling trends in coefficient but does not extend to the inclusion of explanatory variables. This limitation has motivated our current work, which seeks to address these gaps in high-dimensional time series modeling.

We propose a new variation of the modeling that identifies structural breaks and selects relevant variables in high-dimensional settings. As a result, we find a model with fewer parameters to estimate and greater power in identifying these parameters. The novelty we bring is a solution procedure that uses the properties of Fused LASSO iteratively to compare adjacent observations, allowing us to identify structural breaks, identify relevant variables and discard irrelevant ones.

The paper is organized as follows. Section \ref{section:overview} establishes the theoretical foundations of regularized regression, framing these approaches as optimization problems. We propose a dynamic formulation for both the adaptive LASSO and the fused LASSO in Section \ref{section:DynamicLASSO}. Section \ref{section:IFL} introduces our novel algorithm, explaining how it achieves faster and less biased solutions to the proposed problem. Subsection \ref{section:Algorithms} describes its associated algorithms. In Section \ref{section:NumExamples}, we present the results of our experiments. Finally, Section \ref{section:Conclusion} concludes the paper.

\section {Overview of LASSO based methods for sparse regression}
\label{section:overview}

LASSO (Least Absolute Shrinkage and Selection Operator) methods are a class of regularization techniques widely used in statistical modeling and machine learning for shrinkage and variable selection. The standard \textbf{LASSO} \cite{tibshirani1996lasso} introduces an $\ell_1$ penalty on the regression coefficients, encouraging sparsity by shrinking some coefficients exactly to zero, for some \(\lambda \geq 0\), a tuning parameter. The solutions, however, may be inconsistent when there is high correlation among predictor variables and does not possess Oracle properties.

\begin{equation}
    \begin{array}{ll}
        \hat{\beta}^{LASSO} = \argmin\limits_{\beta} & \biggl\{ 
        \mathlarger{\sum}_{i=1}^n \biggl ( y_{i} - \mathlarger{\sum}_{j=1}^p x_{ij} \beta_j  \biggl )^2 
        + \quad \lambda \; \mathlarger{\sum}_{j=1}^p  |\beta_j| \biggl \}\\
    \end{array}
\label{formula:LASSORegressionLagrangian}
\end{equation}

Building on this foundation, the \textbf{adaptive LASSO (adaLASSO)} \cite{Zou2006} improves variable selection consistency by applying data-driven weights \( \omega_j = 1/|\Tilde{\beta_j^\nu}|\) to the $\ell_1$ penalty, allowing differential penalization of coefficients. When $p\geq n$, the Ridge solution can be used as an initial estimate. It is designed to possess Oracle properties under certain regularity conditions. This means that, as the sample size increases, it can be correctly identified as the true model with high probability, leading to better asymptotic performance.

\begin{equation}
    \begin{array}{ll}
        \hat{\beta}^{adaLASSO} = \argmin\limits_{\beta} & \biggl\{ 
        \mathlarger{\sum}_{i=1}^n \biggl ( y_{i} - \mathlarger{\sum}_{j=1}^p x_{ij} \beta_j  \biggl )^2 
        + \quad \lambda  \mathlarger{\sum}_{j=1}^p \omega_j |\beta_j| \biggl \}\\
    \end{array}
\label{formula:AdaLASSORegressionLagrangian}
\end{equation}

The \textbf{fused LASSO} \cite{tibshirani2005fused} extends the LASSO framework further by promoting both sparsity and smoothness through $\ell_1$ penalties on both the coefficients and their successive differences, for some tuning parameters \(\lambda_1 \geq 0\) and \(\lambda_2 \geq 0\), making it especially useful for time series data or ordered features. Together, these methods provide a toolkit for high dimensional structured variable selection and data analysis.

\begin{equation}
    \begin{array}{ll}
        \hat{\beta}^{fusedLASSO} = \argmin\limits_{\beta} & \biggl\{ 
        \mathlarger{\sum}_{i=1}^n \biggl ( y_{i} - \mathlarger{\sum}_{j=1}^p x_{ij} \beta_j  \biggl )^2 
        + \; \lambda_1  \mathlarger{\sum}_{j=1}^p |\beta_j| \; + \; \lambda_2 \mathlarger{\sum}_{j=2}^p |\beta_j - \beta_{j-1}|\biggl \}\\
    \end{array}
\label{formula:FusedLASSORegressionLagrangian}
\end{equation}

The \textbf{generalized LASSO(genLASSO)} \cite{GenLASSO} promotes piecewise constant solutions, encouraging neighboring nodes in a graph to adopt similar or identical values. This results in fused regions where connected nodes share nearly uniform values. The computational framework, based on the dual problem, ensures efficient path computation of the primal optimization problem in its Lagrangian form (\ref{formula:graphGenLASSO}):

\begin{equation}
    \begin{array}{ll}
        \hat{\beta}^{genLASSO} = \argmin\limits_{\beta} & \biggl\{ 
        \mathlarger{\sum}_{i=1}^n \biggl ( y_{i} - \mathlarger{\sum}_{j=1}^p x_{ij} \beta_j  \biggl )^2 
        + \quad \lambda  \mathlarger{\sum}_{(i,j) \in E}^p |\beta_i -\beta_j| + \quad \gamma \lambda \mathlarger{\sum}_{i=1}^p |\beta_i| \biggl \}\\
    \end{array}
\label{formula:graphGenLASSO}
\end{equation}

%Here \(\lambda > 0 \) is a tuning parameter and \( \gamma \in [0,1]\), promote sparsity and encourages 
%neighboring coefficients to be similar and often identical. The computational framework introduces a graph-%based approach leveraging the generalized lasso problem, tailored for graph structures where nodes represent %data points and edges impose penalties reflecting differences between connected nodes.

Where \( \gamma \in [0,1]\). The computational framework introduces a graph-based approach leveraging the generalized lasso problem, tailored for graph structures where nodes represent data points and edges impose penalties reflecting differences between connected nodes.

       % LASSO,  e Ada LASSO
\section {Dynamic formulation of \textbf{adaLASSO} and  \textbf{fused LASSO}}
\label{section:DynamicLASSO}
In the context of time series data, we propose a dynamic formulation for the \textbf{fused LASSO} (\ref{formula:DFLLagrangian}) and the \textbf{fused adaLASSO} (\ref{formula:DFALLagrangian}) to exploit the piecewise-constant structure within a signal and to provide the necessary adjustments for capturing its structural breaks over time.

\begin{equation}
    \begin{array}{lll}
        \hat{\beta}^{FL} =& \argmin\limits_\mathbf{\beta} \biggl\{ 
        \mathlarger{\sum}_{t=1}^T \biggl ( y_t - \mathlarger{\sum}_{j=1}^p x_{t,j} \beta_{t,j}  \biggl )^2 
        &+ \; \lambda_1  \mathlarger{\sum}_{t=1}^T \mathlarger{\sum}_{j=1}^p |\beta_{t,j}| \\ 
        &&+ \; \lambda_2  \mathlarger{\sum}_{t=2}^T \mathlarger{\sum}_{j=1}^p |\beta_{t,j} - \beta_{t-1, j}|\biggl \} \\
    \end{array}
\label{formula:DFLLagrangian}
\end{equation}

\begin{equation}
    \begin{array}{lll}
        \hat{\beta}^{FAL} =& \argmin\limits_\mathbf{\beta} \biggl\{ 
        \mathlarger{\sum}_{t=1}^T \biggl ( y_t - \mathlarger{\sum}_{j=1}^p x_{t,j} \beta_{t,j}  \biggl )^2 
        &+ \; \lambda_1  \mathlarger{\sum}_{t=1}^T \mathlarger{\sum}_{j=1}^p \omega_{\beta,t,j} | \beta_{t,j}| \\ 
        &&+ \; \lambda_2 \mathlarger{\sum}_{t=2}^T \mathlarger{\sum}_{j=1}^p \omega_{\Delta,t,j}|\beta_{t,j} - \beta_{t-1, j}|\biggl \} \\
    \end{array}
\label{formula:DFALLagrangian}
\end{equation}

Where the data is indexed by the time $t$, and \( \lambda_1 \) and \( \lambda_2 \) are tuning parameters, both in (\ref{formula:DFLLagrangian}) and in (\ref{formula:DFALLagrangian}). The first penalty, based on the \(\ell_1\) norm, shrinks the \(\beta_{t,j}\) coefficients towards zero, promoting sparsity. The second penalty leverages the ordered nature of the data, encouraging neighboring \(\beta_{t,j}\) coefficients to be similar and often identical.                  % Fused LASSO e Gen LASSO
\section {Iterative Fused LASSO}
\label{section:IFL}

We propose a novel extension to \textbf{AdaLASSO}, broadening its scope to incorporate relationships among sequentially ordered variables. This extension is termed the \textbf{iterative fused LASSO (IFL)}. The \textbf{IFL} is built upon a three-step procedure designed to solve the \textbf{fused LASSO}, while simultaneously identifying structural breaks and selecting relevant variables. In step $1$ we identify structural breaks, in step $2$ we simplify ("mop") the model, and in step $3$ we select relevant variables.

\subsection{Step 1: identify structural breaks}
\label{section:stp1}
We reformulate the dynamic \textbf{fused AdaLASSO} (\ref{formula:DFALLagrangian}), as a  \textbf{AdaLASSO} (\ref{formula:AdaLASSORegressionLagrangian}), written as:

\begin{equation}
    \begin{array}{ll}
        \hat{b}^{AdaLASSO} =  \argmin\limits_{b} & ( y - X b )^T ( y - X b ) + \lambda \mathlarger \sum_{j=2}^p \omega_{b, j} |{b}| \\
    \end{array}
\label{formula:AdaLASSORegressionLagrangian2}
\end{equation}

Where \( \omega_{b,j} = 1/|\Tilde{b_{b,j}^\nu}|\), we will use $\nu = 1$ for simplicity. We rewrite the problem, so the design matrix $X$ takes the form:

\be
X =
\begin{bmatrix}
diag(x_1)  | \cdots | diag(x_p)
\end{bmatrix} = 
\begin{bmatrix}
    \begin{bmatrix}
        \begin{array}{ccc}
            x_{11}  &\cdots    &0       \\
            \vdots 	&\ddots    &\vdots  \\
            0       &\cdots    &x_{1n}  \\
        \end{array}
    \end{bmatrix}
    \cdots
    \begin{bmatrix}
        \begin{array}{ccc}
            x_{p1}  &\cdots    &0       \\
            \vdots 	&\ddots    &\vdots  \\
            0       &\cdots    &x_{pn}  \\
        \end{array}
    \end{bmatrix}
\end{bmatrix}
\label{formula:Xcross}
\ee

and the vector $b$ is: 
\be
b = vec(B) = 
\begin{bmatrix}
\begin{array}{c}
\beta_{(1)}\\
\beta_{(2)}\\
\vdots\\
\beta_{(p)}
\end{array}
\end{bmatrix}
,\quad
B = 
\begin{bmatrix}
\begin{array}{ccc}
\beta_{11}  &\cdots       &\beta_{1n} \\
\vdots        &\ddots       &\vdots\\
\beta_{p1}  &\cdots       &\beta_{pn}
\end{array}
\end{bmatrix}
%=
%\begin{bmatrix}
%\begin{array}{c}
%\beta^T_1 \\
%\vdots \\
%\beta^T_n  
%\end{array}
%\end{bmatrix}
%= 
%\begin{bmatrix}
%\beta_{(1)} \cdots \beta_{(p)}
%\end{bmatrix}
\label{formula:vectorb}
\ee

In this context, $X$ is an $n \times np$ matrix, and $b$ is $np \times 1$ vector. Our goal is to detect recurring values in each of the $p$ components at observation $t$ of $\beta_{t,j}$, $t = 2, \cdots, T$, $j = 1, \cdots, p$. To achieve this, we need to define supporting matrices that will help us manage the lag differences between observations and the number of recurring values in each component $p$. Denoting the difference matrix $L_d^{(k)}$ for component $p$ as follows: \(d\) represents the lag difference between observations, and \(k\) represents the number of distinct values in feature $p$. It allows us to define the difference matrix, $L_1$.

\[
L_d^{(k)} = I_k - 
\begin{bmatrix}
\begin{array}{ll}
0_{d,(k-d)}  &0_{d,d} \\
I_{(k-d)}    &0_{(k-d),d}
\end{array}
\end{bmatrix}, 
\quad
L_1 =  
\begin{bmatrix}
\begin{array}{ccc}
L_1^{(k)} &0       &0\\
0           &\ddots  &0\\
0           &0       &L_1^{(k)}
\end{array}
\end{bmatrix}
\]

$L_1$ is a diagonal matrix that shows the number of distinct values in each one of the $p$ components at one lag difference. This allows us to solve the structural break problem:

\be
\hat{\theta}^{SB} =  \argmin\limits_{\theta} ( y - XL_1^{-1}\theta )^T ( y - XL_1^{-1}\theta ) + \lambda \mathlarger \sum_{i=2}^p \omega_{\theta, i} |{\theta_i}|
\label{formula:AdaLASSOTheta1}
\ee

where $L_1^{-1}\theta = b$, such that $\theta = L_1 b \rightarrow b = L_1^{-1}\theta$. At this stage, we can assume that each of the $p$ features has distinct values with no repetitions. Given $n$ observations and $p$ features, we can express the following:

\be
L_1 =  
\begin{bmatrix}
\begin{array}{ccc}
L_1^{(n)} &0       &0\\
0           &\ddots  &0\\
0           &0       &L_1^{(n)}
\end{array}
\end{bmatrix}
, \; \text{and} \; 
\theta =
\begin{bmatrix}
\begin{array}{c}
\beta_{1,1}\\
\beta_{2,1} - \beta_{1,1}\\
\cdots\\
\beta_{n,1} - \beta_{(n-1),1}\\
\cdots\\
%\hline
\beta_{1, p}\\
\beta_{2,p} - \beta_{1,p}\\
\cdots\\
\beta_{n, p} - \beta_{(n-1), p}\\
\end{array}
\end{bmatrix}
\label{formula:theta}
\ee

By solving (\ref{formula:AdaLASSOTheta1}), we obtain an estimate $\hat{\theta}^{SB}$. The best tuning parameter $\lambda$ is selected using the Bayesian Information Criterion (BIC). The Ada LASSO solution estimates which components of $\beta_{t, j} - \beta_{t-1, j}$ can be set to zero. Components that are non-zero indicate differences between consecutive values, representing a structural break in the time series at observation $t$ for the $j$-th component. However, the Structural Break solution does not estimate the specific values of $\beta_{t, j}$. Instead, it identifies the observations where these values differ, thus indicating where a structural break occurs in the data.

\subsection{Step 2: simplify ("mop") the model}
\label{section:stp2}
We propose a procedure that rewrites the problem in a reduced form that takes into account the repeated values of $\beta_{t,j}$ and the breaks found in the $\hat{\theta^{SB}}$ solution of (\ref{formula:AdaLASSOTheta1}). First, we must keep track of all recurring values in each $p$ component. To do this we define $\beta_{in}$, a $p \times 1$ vector, that tracks the number of distinct values at each one of the $p$ components. At the first iteration $\beta_{in} = [n, \cdots, n]$, assuming all $\beta_{t,j}$, $t = 1, \cdots, n$ and $j = 1, \cdots, p$ are distinct.

Next, we define $\gamma_d$, a $np \times 1$ vector that accumulates the number of different values of $\beta_{t,j}$ at each $j$ component, $j = 1 \cdots, p$. After identifying the number of different values of $\beta_{t,j}$, we define $\beta_{out}$, a $p \times 1$ vector, that tracks the number of distinct values at each one of the $p$ components. If no repeated values were identified, then $\beta_{out} = \beta_{in}$. Therefore each component of $\beta_{out}$ is at most as high as $\beta_{in}$. 

Then we define $n_{\beta_{in}}$, the number of different values in $\beta_{in}$ as the sum of its components, and also $n_{\beta_{out}}$, the number of different values in $\beta_{out}$ as the sum of its components. We define a matrix $M$ that has as many lines as the number of different values in $\beta_{in}$, and as many columns as the number of different values in $\beta_{out}$. We build matrix $M$ in such a way that it maps the solution $\hat{b}$ as:

\[
b = \textbf{M}\, b_{reduced}
\]

\subsection{Step 3: select relevant variables}
\label{section:stp3}
We are now able to solve a reduced problem:

\be
    \begin{array}{ll}
        \hat{\gamma}^{AdaLASSO} = \argmin\limits_{\gamma} & ( y - H\gamma)^T ( y - H\gamma ) + \lambda \mathlarger \sum_{i=1}^{K} \omega_i |{\gamma_i}|
    \end{array}
    \label{formula:AdaLASSOreduced}
\ee

where $H = [X\,M] $ and $\gamma = b_{reduced}$ ($K \times 1$), and $K$ is the number of distinct features in the reduced model. The reduced problem estimates $\gamma$ such that it takes into account the structural breaks identified in (\ref{formula:AdaLASSOTheta1})  and also produces estimates for $\beta$ that solves (\ref{formula:AdaLASSORegressionLagrangian2}), and estimates: 

\[
\hat{b} = M \; \hat{\gamma}
\]

Therefore, our proposed procedure estimates $\beta_{t,j}$, enabling the identification of structural breaks, and proper selection of relevant variables.
    
\subsection{Algorithms used for \textit{moping} the model}
\label{section:Algorithms}

At the end of stage (\ref{section:stp2}), it is possible to identify elements of $\beta_{t,j}$ that remain constant over time. In the next stage, these repeated values can be estimated jointly, acknowledging their equality. To facilitate this, we construct a vector, denoted $\gamma_d$, which accumulates the number of distinct values of $\beta_{t,j}$ across time for each component $j$.

Algorithm \ref{algorithm:gamma_d} describes the construction of $\gamma_d$, while Algorithm \ref{algorithm:matrixM} outlines how the matrix $M$ is formed. Let $n_{\beta_{in}}$ denote the total number of distinct components in $\beta_{in}$, and $n_{\beta_{out}}$ the total in $\beta_{out}$. The matrix $M$ maps the reduced vector $b_{\text{reduced}}$ to the full $\beta_{t,j}$, such that it consists of columns containing only $0$s and $1$s that correctly position the elements of $b_{\text{reduced}}$ via pre-multiplication.

\begin{comment}
\begin{algorithm}
    \caption{Vector $\gamma_d$. Inputs: ($\beta$, $n$,$p$, $\beta_{in}$, $d$. Output: vector $\gamma_d$}
    \begin{algorithmic}%[1] % for counting line numbers
        \label{algorithm:gamma_d}
        \STATE Initialize $\gamma_{d_{boolean}} \leftarrow$ abs($\beta) > 0$
        \STATE Initialize $\gamma_d \leftarrow \mathbf{0}_{n_{\beta_{in}} \times 1}$
        \FOR{$i \leftarrow 0$ to $(n-1)$}
           \FOR{$j \leftarrow 1$ to $p$}
               \IF {$j \leq d$}
                   \STATE $\gamma_d[i+j]  = 1$
               \ELSIF{$\gamma_d[i+j]  =$ TRUE}
                   \STATE $\gamma_d[i+j]  = \gamma_d[i+j-1] + 1$
               \ENDIF
           \ENDFOR
        \ENDFOR
        \STATE Returns $\gamma_d$  
    \end{algorithmic}
\end{algorithm}

\begin{algorithm}
    \caption{Selection Matrix M. Inputs: ($\beta$, $\beta_{in}$, $\beta_{out}$, $\gamma_d$), Output: matrix M}
    \begin{algorithmic}%[1] % for counting line numbers
        \label{algorithm:matrixM}
        \STATE Initialize $M \leftarrow \mathbf{0}_{n_{\beta_{in}} \times n_{\beta_{out}} }$
        \FOR{$i \leftarrow 1$ to $n_{\beta_{in}}$}
            \IF {$\gamma_d \neq 0$}
                \STATE $M(i, \gamma_d[i]) \leftarrow 1$
            \ELSE
                \STATE $M(i, \gamma_d[i-d]) \leftarrow 1$
            \ENDIF
        \ENDFOR
        \STATE Returns $M$  
    \end{algorithmic}
\end{algorithm}
\end{comment}

\begin{minipage}[t]{0.45\textwidth}
    \begin{algorithm}[H]
        \caption{Vector $\gamma_d$. \\ Inputs: ($\beta$, $n$, $p$, $\beta_{in}$, $d$).\\ Output: vector $\gamma_d$}
        \begin{algorithmic}
            \label{algorithm:gamma_d}
            \STATE Initialize $\gamma_{d_{\text{boolean}}} \leftarrow$ abs($\beta$) $> 0$
            \STATE Initialize $\gamma_d \leftarrow \mathbf{0}_{n_{\beta_{in}} \times 1}$
            \FOR{$i \leftarrow 0$ to $(n-1)$}
                \FOR{$j \leftarrow 1$ to $p$}
                    \IF{$j \leq d$}
                        \STATE $\gamma_d[i+j]  \leftarrow 1$
                    \ELSIF{$\gamma_d[i+j]  =$ TRUE}
                        \STATE $\gamma_d[i+j]  \leftarrow \gamma_d[i+j-1] + 1$
                    \ENDIF
                \ENDFOR
            \ENDFOR
            \STATE \textbf{Return} $\gamma_d$
        \end{algorithmic}
    \end{algorithm}
\end{minipage}
\hfill
\begin{minipage}[t]{0.45\textwidth}
    \begin{algorithm}[H]
        \caption{Selection Matrix M. \\Inputs: ($\beta$, $\beta_{in}$, $\beta_{out}$, $\gamma_d$). \\Output: matrix $M$}
        \begin{algorithmic}
            \label{algorithm:matrixM}
            \STATE Initialize $M \leftarrow \mathbf{0}_{n_{\beta_{in}} \times n_{\beta_{out}}}$
            \FOR{$i \leftarrow 1$ to $n_{\beta_{in}}$}
                \IF{$\gamma_d[i] \neq 0$}
                    \STATE $M(i, \gamma_d[i]) \leftarrow 1$
                \ELSE
                    \STATE $M(i, \gamma_d[i-d]) \leftarrow 1$
                \ENDIF
            \ENDFOR
            \STATE \textbf{Return} $M$
        \end{algorithmic}
    \end{algorithm}
\end{minipage}
          % Fused LASSO, e Iterative Fused LASSO
\section{Numerical examples}
\label{section:NumExamples}
To evaluate the performance of our procedure, we applied the \textbf{IFL} algorithm as described in Section \ref{section:IFL} in $R$, using the \textbf{glmnet} package. As benchmarks, we used the \textbf{genLASSO} formulation (\ref{formula:graphGenLASSO}), also in $R$. 

\subsection{Monte Carlo simulation results}
\label{subsection:MCexperiment}
The simulated experiments involved generating 18 sets of 100 time series each. For each scenario, we constructed a four-regime sample using distinct sets of design matrices. The experiments varied the total number of features (\(p\)) and the number of relevant, non-zero features (\(q\)). Additionally, we examined the effect of varying the number of observations within each regime (\(n/R\)). The performances of the  \textbf{IFL} and  \textbf{genLASSO} solutions were compared for all scenarios, with the Oracle solutions serving as benchmarks.

The simulated scenarios included varying the number of observations per regime ($n/R = 30$, $n/R = 50$), the total number of features ($p = 20$, $p = 30$, $p = 40$), and the total number of relevant (non-zero) features ($q = 2$, $q = 5$, $q = 10$).

\begin{figure}[!ht]
    \centering
    \begin{minipage}{0.48\textwidth}
        \centering
        \includegraphics[width=\textwidth]{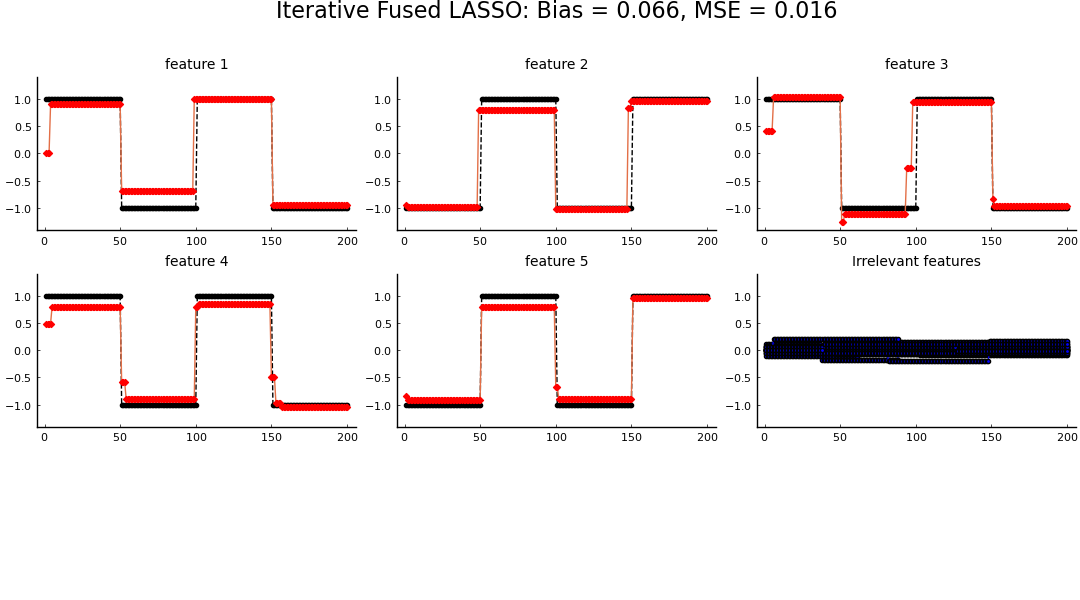}
        \caption{IFL solution example}
        \label{image:IFL_Example}
    \end{minipage}%
    \hfill
    \begin{minipage}{0.48\textwidth}
        \centering
        \includegraphics[width=\textwidth]{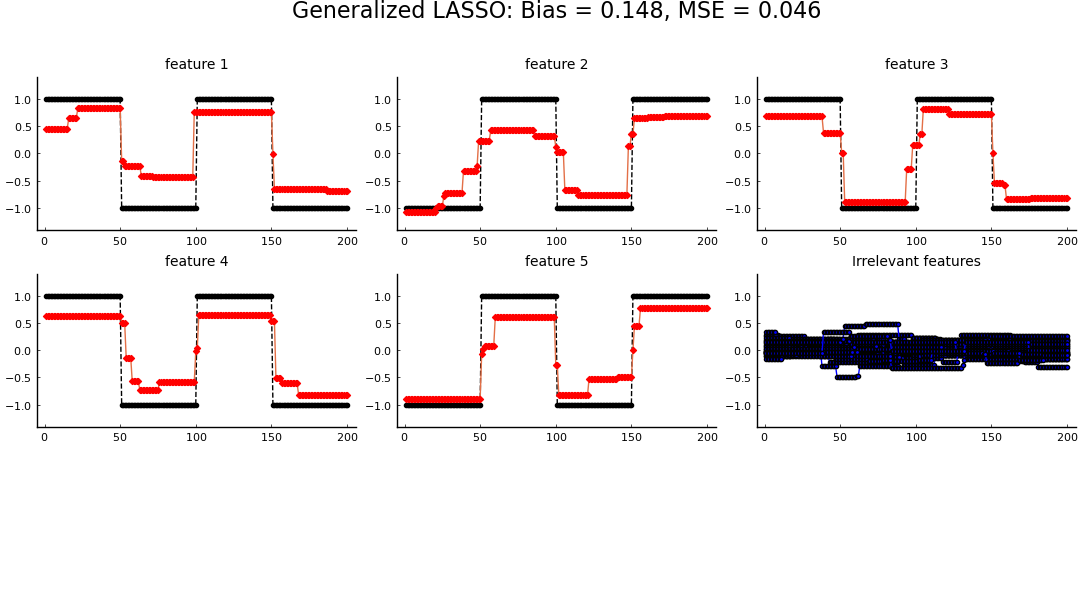}
        \caption{GenLASSO solution example}
        \label{image:GLASSO_Example}
    \end{minipage}
    %\caption{Comparison of solutions: Iterative Fused LASSO vs. Generalized LASSO}
    \label{fig:comparison_solutions}
\end{figure}

Figures \ref{image:IFL_Example} and \ref{image:GLASSO_Example} illustrate examples of the estimated feature values over time for one instance solved using each algorithm. This instance involved \(n=200\) observations, \(p=30\) features, and \(q=5\) non-zero features. In both cases, \textbf{IFL} demonstrated smaller bias and mean squared error (MSE) compared to \textbf{genLASSO}. Notably, it generated solutions that were less noisy and more closely aligned with the synthesized values.

Tables~\ref{Table:MCIFL} and~\ref{Table:MCGenLASSO} present the results for \textbf{IFL} and \textbf{genLASSO}, respectively. They summarize the average bias and MSE across the Monte Carlo scenarios.
Comparing both procedures across scenarios reveals that for less sparse problems ($q = 10$), \textbf{genLASSO} exhibits lower bias and MSE than \textbf{IFL}. Conversely, for sparser problems ($q = 2$), \textbf{genLASSO} shows higher bias and MSE compared to \textbf{IFL}. As the number of relevant variables decreases,  \textbf{IFL} consistently provides solutions with lower bias than  \textbf{genLASSO}.

\begin{table}[!ht]
\centering
%\caption{Iterative Fused LASSO (Esquerda) e Generalized LASSO (Direita)}
%\label{Table:Comparison}
\begin{minipage}{0.48\linewidth}
    \centering
    \caption{Iterative Fused LASSO}
    \label{Table:MCIFL}
    \begin{tabular}{ccc}
    \toprule \toprule
        & \underline{$n/R=30$} & \underline{$n/R=50$} \\
        \underline{$q$\textbackslash$p$} &
        \underline{\begin{tabular}{ccc} 20 & 30 & 40 \end{tabular}} &
        \underline{\begin{tabular}{ccc} 20 & 30 & 40 \end{tabular}} \\
        \\ \multicolumn{3}{c}{\textbf{Panel (a): Bias}} \\
        \begin{tabular}{c}
            2 \\ 5 \\ 10 \\
        \end{tabular} &
        \begin{tabular}{lll}
            0.07 & 0.08 & 0.09 \\ 0.20 & 0.24 & 0.27 \\ 0.54 & 0.49 & 0.45 \\
        \end{tabular} &
        \begin{tabular}{lll}
            0.04 & 0.03 & 0.04 \\ 0.08 & 0.07 & 0.08 \\ 0.22 & 0.34 & 0.38 \\
        \end{tabular} \\
        \\ \multicolumn{3}{c}{\textbf{Panel (b): MSE}} \\
        \begin{tabular}{c}
            2 \\ 5 \\ 10 \\
        \end{tabular} &
        \begin{tabular}{lll}
            0.02 & 0.02 & 0.02 \\ 0.12 & 0.15 & 0.17 \\ 0.51 & 0.43 & 0.37 \\
        \end{tabular} &
        \begin{tabular}{lll}
            0.01 & 0.01 & 0.01 \\ 0.03 & 0.02 & 0.03 \\ 0.16 & 0.27 & 0.29 \\
        \end{tabular} \\
        \bottomrule
    \end{tabular}
\end{minipage}%
\hfill
\begin{minipage}{0.48\linewidth}
    \centering
    \caption{Generalized LASSO}
    \label{Table:MCGenLASSO}
    \begin{tabular}{ccc}
    \toprule \toprule
        & \underline{$n/R=30$} & \underline{$n/R=50$} \\
        \underline{$q$\textbackslash$p$} &
        \underline{\begin{tabular}{ccc} 20 & 30 & 40 \end{tabular}} &
        \underline{\begin{tabular}{ccc} 20 & 30 & 40 \end{tabular}} \\
        \\ \multicolumn{3}{c}{\textbf{Panel (a): Bias}} \\
        \begin{tabular}{c}
            2 \\ 5 \\ 10 \\
        \end{tabular} &
        \begin{tabular}{lll}
            0.13 & 0.11 & 0.10 \\ 0.22 & 0.20 & 0.20 \\ 0.37 & 0.35 & 0.35 \\
        \end{tabular} &
        \begin{tabular}{lll}
            0.12 & 0.10 & 0.09 \\ 0.17 & 0.14 & 0.14 \\ 0.27 & 0.24 & 0.23 \\
        \end{tabular} \\
        \\ \multicolumn{3}{c}{\textbf{Panel (b): MSE}} \\
        \begin{tabular}{c}
            2 \\ 5 \\ 10 \\
        \end{tabular} &
        \begin{tabular}{lll}
            0.04 & 0.03 & 0.02 \\ 0.10 & 0.09 & 0.08 \\ 0.25 & 0.23 & 0.23 \\
        \end{tabular} &
        \begin{tabular}{lll}
            0.03 & 0.02 & 0.02 \\ 0.06 & 0.05 & 0.04 \\ 0.14 & 0.12 & 0.11 \\
        \end{tabular} \\
        \bottomrule
    \end{tabular}
\end{minipage}
\end{table}

\begin{figure}[ht]
\centering
\includegraphics[scale=0.39]{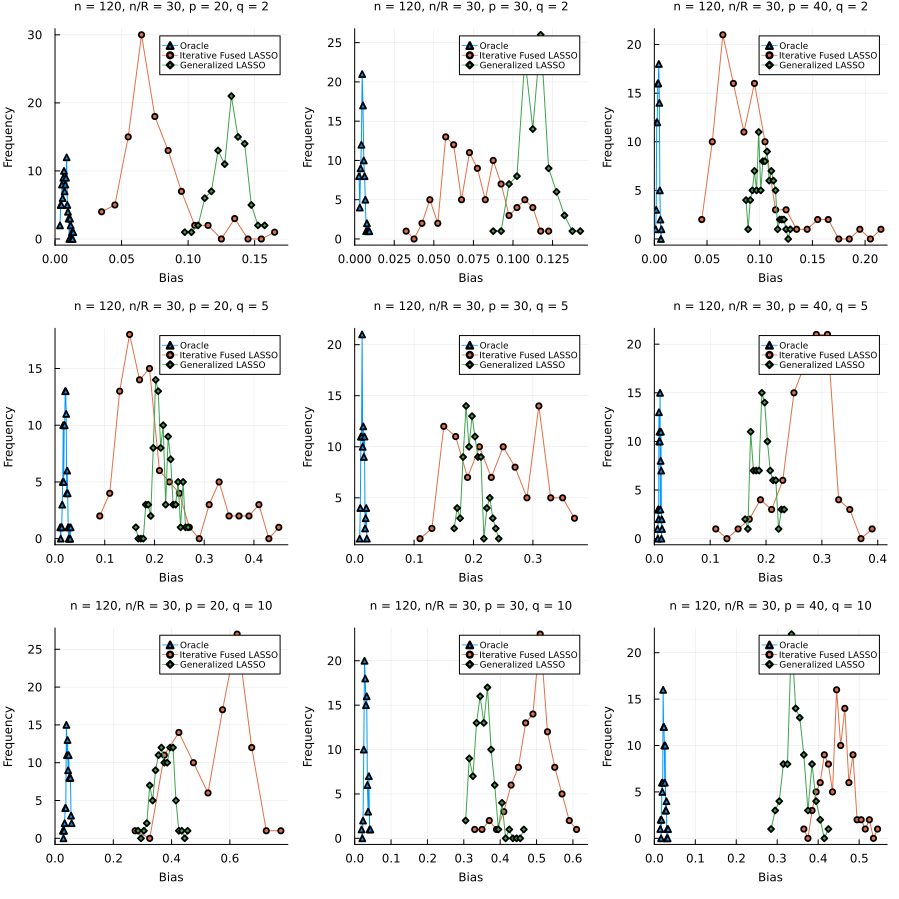}
\caption{Distribution of the bias of the Iterative Fused Lasso, Generalized LASSO and Oracle estimators for the parameters over 100 Monte Carlo replications. Different combinations of candidate and relevant variables. The sample size equals 120 observations.}
\label{image:Hist_nR30_Bias}  
\end{figure}

\begin{figure}[!ht]
\centering
\includegraphics[scale=0.39]{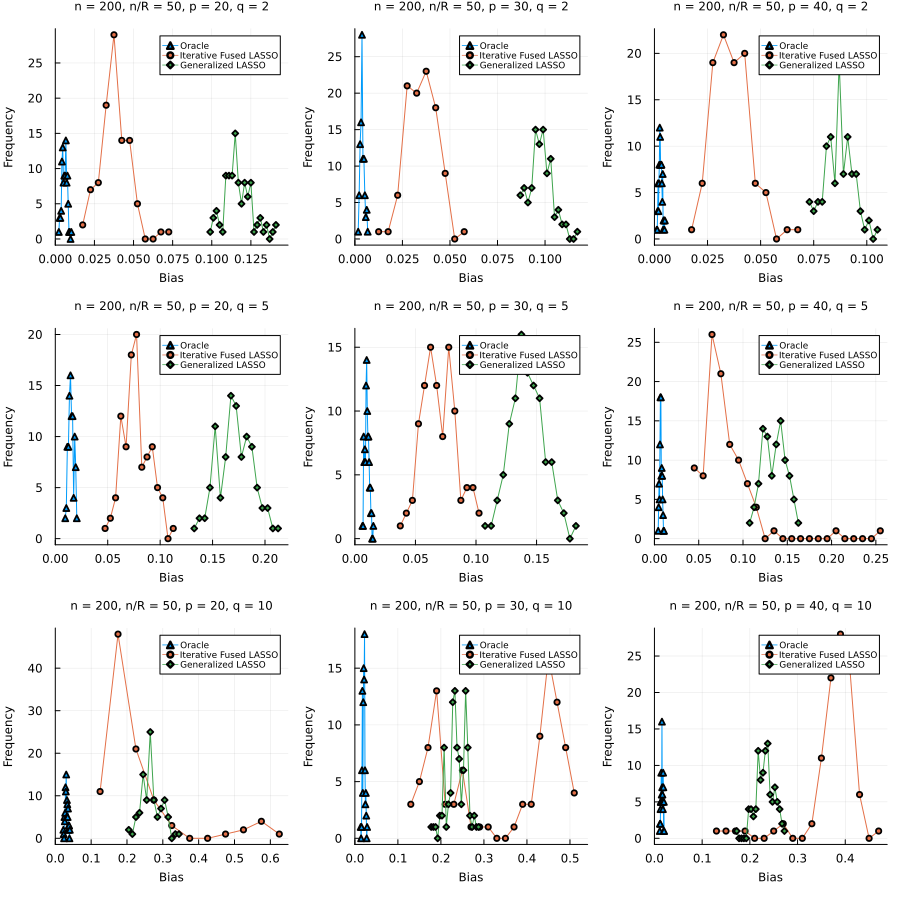}
\caption{Distribution of the bias of the Iterative Fused Lasso, Generalized LASSO and Oracle estimators for the parameters over 100 Monte Carlo replications. Different combinations of candidate and relevant variables. The sample size equals 200 observations.}
\label{image:Hist_nR50_Bias}  
\end{figure}

Further insights into the performance differences are illustrated in Figures \ref{image:Hist_nR30_Bias} and \ref{image:Hist_nR50_Bias}, which display the distribution of the bias for both algorithms alongside the Oracle solution for $n = 120$ ($n/R = 30$) observations, and $n = 200$ ($n/R = 50$) observations, with different combinations of candidate and relevant variables. These figures demonstrate that as the number of relevant variables increases, both \textbf{IFL} and \textbf{genLASSO} tend to deviate more from the Oracle. However, \textbf{IFL} generally remains closer to the Oracle compared to \textbf{genLASSO}.

A critical observation pertains to the convergence time of each method. Across all solved cases, the \textbf{IFL} algorithm converges in approximately $10\%$ of the time required by \textbf{genLASSO}. This significant difference renders \textbf{genLASSO} impractical for larger problems, such as those involving $n/R > 50$ observations.

\subsection{Real-World Dataset results}
\label{subsection:RWDataset}

For the "real-world" dataset, we solve the problem of an active fund return estimate as a dynamic fused AdaLASSO (\ref{formula:DFALLagrangian}). 

\be
 r_{0,t} = \sum_{i=1}^K N_{i,t} \frac{P_{i,t}}{Q_{t-1}} - 1
\label{formula:Qreturn3}
\ee

where $r_{0,t}$ is the arithmetic return of the fund, $P_{i,t}$ is the price of one share of asset $i$ at time $t$, while $N_{i,t}$ represents the number of shares, $Q_{t-1}$ is the value of the fund at time $t-1$. Our goal is to estimate the observed value of the active fund at time $t$, denoted $y_t$, where $x_{t,j}$ represent the individual return of each asset, $\beta_t$ represent the number of shares, and $\epsilon_t$ represent measure disturbances. In matrix form: 
\[
y_t = r_{0,t} + 1, \quad y_t = x_t^T \beta_t + \epsilon_t, \quad y = X \, b + \epsilon 
\]
\[
x_t = \left [ \frac{P_{i_1,t}}{Q_{t-1}}, \cdots, \frac{P_{i_K,t}}{Q_{t-1}} \right ]^T, \;
\beta_t = \left [ N_{i_1,t}, \cdots, N_{i_K,t} \right ], \; b = \begin{bmatrix}
\begin{array}{c}
\beta_{(1)}\\
\beta_{(2)}\\
\vdots\\
\beta_{(K)}
\end{array}
\end{bmatrix}
\]

We estimate the number of shares in time from a synthetic portfolio using $p=20$ series of daily stock returns, where ($q=3$) relevant non-zero features were active at certain points, resulting in $2$ regimes and $4$ structural breaks. Each regime comprised $758$ observations, spanning from January 2, 2021, to December 23, 2023. To construct this portfolio of returns, we utilized quotes from B3 (\textit{Brasil Bolsa Balcao}), Brazil's main hub for trading equities, derivatives, and other financial instruments. The mix of relevant stocks in the portfolio was altered at a single point in time. Initially, the portfolio comprised $25\%$ of stocks from $VALE3$, $25\%$ of stocks from $PETR4$, and $50\%$ of stocks from $ELET6$. After the change, the portfolio shifted to $75\%$ of stocks from $VALE3$, $0\%$ of stocks from $PETR4$, and $25\%$ of stocks from $ELET6$. All other quotes remained non-relevant to the portfolio; however, their returns influenced the observed data in the sense that we observed the portfolio returns.

\begin{comment}
Below is a breakdown of the relevant companies and their respective sectors:

\begin{itemize}
    \item{VALE3 (Vale S.A.): returns as $r_{t, 1}$} One of the world's largest mining companies, specializing in iron ore and nickel.
    \item{PETR4 (Petrobras): returns as $r_{t, 2}$} One of the world's largest mining companies, specializing in iron ore and nickel.
    \item{ELET6 (Eletrobras): returns as $r_{t, 3}$} One of the world's largest mining companies, specializing in iron ore and nickel.
    %\item{ITUB4 (Itaú Unibanco): returns as $r_{t, 4}$} One of the world's largest mining companies, specializing in iron ore and nickel.
    %\item{BBDC4 (Banco Bradesco): returns as $r_{t, 5}$} One of the world's largest mining companies, specializing in iron ore and nickel.
    %\item{ABEV3 (Ambev): returns as $r_{t, 6}$} A leading beverage company in Latin America, producing beer and soft drinks.     
\end{itemize}
\end{comment}

%Simulated case studies and a real-world dataset case of a synthetic portfolio of stocks are included to demonstrate the %methodology's applicability and robustness.
%VALE3, PETR4, ELET6, ITUB4, BBDC4, ABEV3
%beta1 <- c(c(0.25, 0.75), rep(0, p-5))
%beta2 <- c(c( 0.6, 0.40), rep(0, p-5))

\begin{figure}[htbp]
    \centering
    \begin{minipage}{0.48\textwidth}
        \centering
        \includegraphics[width=\textwidth]{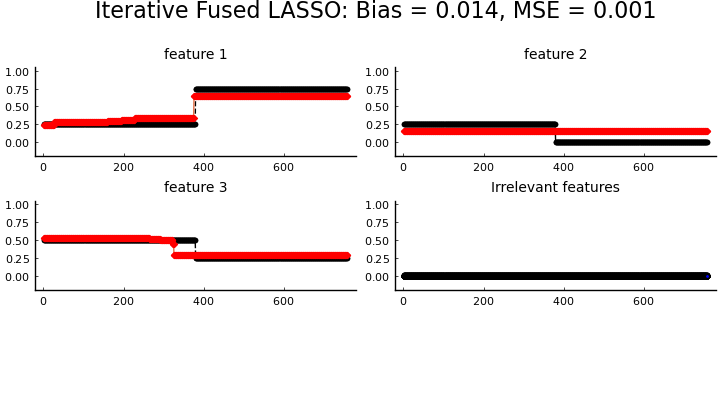}
        \caption{IFL: portfolio dataset}
        \label{image:RWD_IFL}
    \end{minipage}%
    \hfill
    \begin{minipage}{0.48\textwidth}
        \centering
        \includegraphics[width=\textwidth]{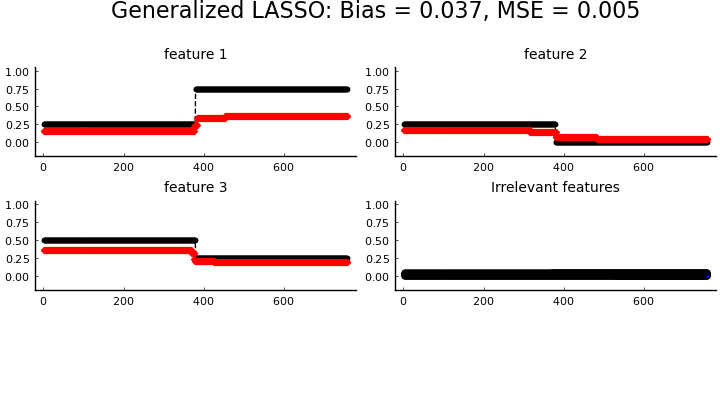}
        \caption{GenLasso: portfolio dataset}
        \label{image:RWD_GL}
    \end{minipage}
    %\caption{Comparison of solutions for the portfolio dataset: IFL vs. Generalized LASSO}
    \label{fig:RWD_Comparison}
\end{figure}

Figures \ref{image:RWD_IFL} and \ref{image:RWD_GL} illustrate the estimated feature values over time for each algorithm, respectively. When applied to noisy real data from stock returns, neither algorithm was able to perfectly track every feature value. However, both algorithms achieved relatively small bias and MSE values.            % Simulação MC (1,800 series) + Exemplo Portfolio Ações
\section {Conclusion}
\label{section:Conclusion}
Our proposed methodology effectively addresses the challenges of variable selection and structural break detection in high-dimensional time series analysis. The integration of regularization techniques like \textbf{fused LASSO} and \textbf{adaLASSO}, coupled with our efficient algorithm, demonstrates superior performance in both simulated and real-world scenarios. In all replications of our Monte Carlo experiments the \textbf{IFL} algorithm was able to correctly identify structural breaks, and select relevant variables. This work not only advances the state of high-dimensional time series modeling but also provides a robust framework with practical applicability, paving the way for its adoption in diverse domains requiring sophisticated analytical tools. 

% ---- Bibliography ----
\bibliographystyle{splncs04}
\bibliography{references} % The name of your .bib file (without .bib extension)

\end{document}